\documentclass[runningheads]{llncs}
\usepackage{graphicx}
\usepackage{amssymb}
\usepackage{multirow}
\usepackage{xcolor}
\usepackage{hyperref}

\begin{document}
\title{Recursive Refinement Network for Deformable Lung Registration between Exhale and Inhale CT Scans}%
\author{Xinzi He\inst{1} \and
Jia Guo\inst{3} \and
Xuzhe Zhang \inst{1} \and
Hanwen Bi \inst{1}\and
Sarah Gerard \inst{2}\and
David Kaczka \inst{2}\and
Amin Motahari \inst{2}\and
Eric Hoffman \inst{2}\and
Joseph Reinhardt  \inst{2}\and
R. Graham Barr  \inst{3}\and
Elsa Angelini \inst{1}\and
Andrew Laine \inst{1}}
\authorrunning{F. Author et al.}
%
\institute{Columbia University, New York, NY, 10025, USA \and
University of Iowa, Iowa City, Iowa, 52242, USA \and
Columbia University Medical Center, New York, NY, 10025, USA 
}
\maketitle              
\begin{abstract}
Unsupervised learning-based medical image registration approaches have witnessed rapid development in recent years. We propose to revisit a commonly ignored while simple and well-established principle: recursive refinement of deformation vector fields across scales. We introduce a recursive refinement network (RRN) for unsupervised medical image registration, to extract multi-scale features, construct normalized local cost correlation volume and recursively refine volumetric deformation vector fields. RRN achieves state of the art performance for 3D registration of expiratory-inspiratory pairs of CT lung scans. On DirLab COPDGene dataset, RRN returns an average Target Registration Error (TRE) of 0.83 mm, which corresponds to a 13\% error reduction from the best result presented in the leaderboard \footnote{https://www.dir-lab.com/Results.html}. In addition to comparison with conventional methods, RRN leads to 89\% error reduction compared to deep-learning-based peer approaches. Code: \href{https://github.com/Novestars/Recursive_Refinement_Network}{https://github.com/Novestars/Recursive\_Refinement\_Network}.

\end{abstract}
\section{Introduction}
\begin{figure}[t]
\includegraphics[width=\textwidth]{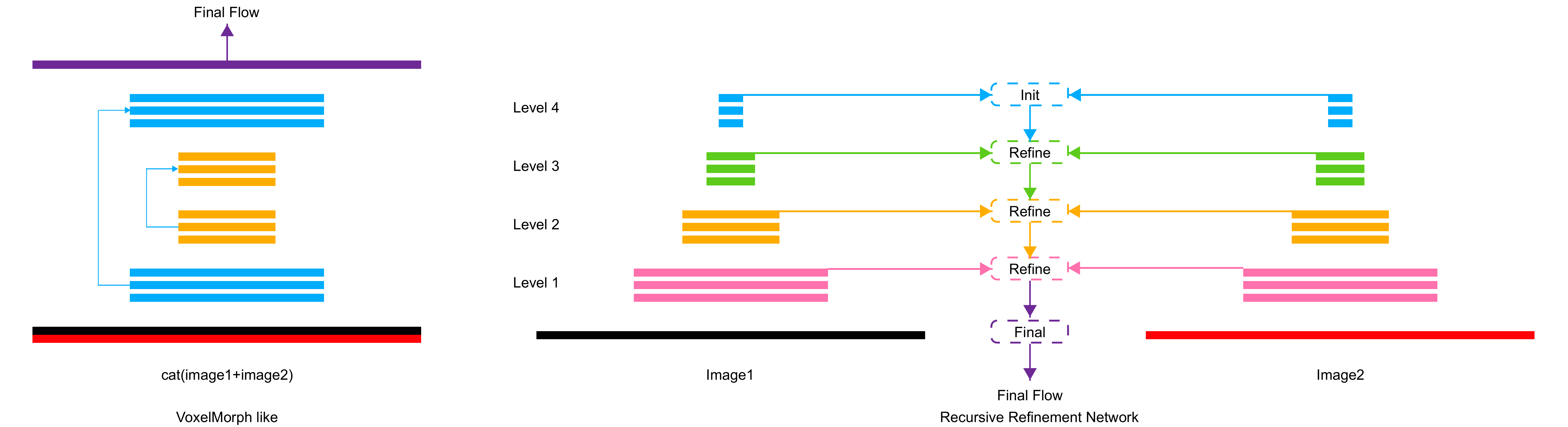}
\caption{VoxelMorph \cite{balakrishnanVoxelMorphLearningFramework2019} approach vs. our proposed Recursive Refinement Network.} \label{fig1}
\end{figure}

Deformable registration of lung images has important implications for the matching of CT scans acquired at different time points or lung volumes to an appropriate reference frame.  Thus, it may have clinical utility for planning therapeutic interventions (i.e., external beam radiation for cancer treatment), or providing unique insight into derangements of regional parenchymal deformation and mechanics that are otherwise unapparent.  For example, the deformation field produced by image registration can be analyzed to measure local biomechanical properties of the lung \cite{amelonThreedimensionalCharacterizationRegional2011,bodduluriBiomechanicalCTMetrics2017,kaczkaAnalysisRegionalMechanics2011}. Registration-derived estimates of local lung volume change have also been shown to correlate with regional ventilation measured by Xenon-enhanced CT imaging \cite{reinhardtRegistrationbasedEstimatesLocal2008}.  Similarly, dynamic CT imaging has been used in cancer patients for optimal beam localization during radiation therapy. However, lung registration remains a challenging task in image processing, given a priori assumptions on the elasticity of lung tissue deformation, nonlinear deformations that occur over large volume excursions, as well as computational complexity and non-convexity from the perspective of optimization. 

VoxelMorph \cite{balakrishnanVoxelMorphLearningFramework2019} pioneered the use of convolutional neural networks (CNNs) for unsupervised deformable image registration relying on a encoder-decoder structures with skip connections. Since VoxelMorph registration accuracy remained poor in some applications, later works focused on designing more powerful CNN architectures. Recursive cascaded network (RCN) \cite{zhaoRecursiveCascadedNetworks2019} outperformed VoxelMorph in various tasks including liver registration and brain MRI template registration. The main idea was to use stacked CNN to recursively predict velocity fields instead of the global deformable vector fields at once such as VoxelMorph. However, one drawbacks of such cascaded scheme is the linear increase in the number of parameters. Also, RCN requires pre-training of a subnetwork first, resulting in a complex training procedure in practice.

Recently, many lightweight networks have been proposed and achieved competitive accuracy, such as the Laplacian pyramid Network (LIPNet) \cite{mokLargeDeformationDiffeomorphic2020} and the Dual-Stream Pyramid Network (DSPN)\cite{huDualStreamPyramidRegistration2019}. LIPNet performs  coarse-to-fine estimation of the deformation field by iterative updates across levels of a Laplacian pyramid using networks with individually trained weights.  It demonstrated superiority over VoxelMorph with less parameters compared to RCN. DSPN further reduced the number of weights by replacing image pyramids with feature pyramids. However, these methods significantly deteriorate when handling large displacements, due to very simple flow field estimators and limited abilities to capture long-range image similarities \cite{devosDeepLearningFramework2018} \cite{fuDeepLearningMedical2020}. 
To solve these issues, we propose a Recursive Refinement Network (RRN) and verify the performance on a changeling dataset, the DirLab COPDGene \cite{castilloReferenceDatasetDeformable2013} \cite{castilloFrameworkEvaluationDeformable2009}. 
%
RRN contains 3 main components: (1) a pyramidal feature extractor from the two input images; (2) a normalization and local cost correlation layer that allows constructing normalized cost correlation volumes in a memory efficient manner; (3) multiple lightweight deformation field estimators which generate initial deformation vector fields (DVFs) or refine the DVFs across scales (levels) of the pyramid of features. 

\section{Methods}

\subsection{Recursive Refinement Network (RRN)}
Figure \ref{fig1} illustrates the proposed RRN architecture of our network, compared to VoxelMorph \cite{balakrishnanVoxelMorphLearningFramework2019}. First, we replace image pyramids, which have been widely used in traditional methods, by feature pyramids generated by CNNs, as CNN features are more robust to image variability. Second, we introduce normalized local cost correlation volumes for deformation field prediction, assuming they represent image registration better than warped features. Third, an recursive refinement scheme is used that gathers deformations at each level of the feature pyramid  to generate more accurate high-resolution deformation vector fields. 

In the following sections, we detail these components individually, including the feature pyramid extractor, feature spatial transformation, local cost correlation volume, cost correlation volume normalization and DVF estimator. 

\subsubsection{Feature Pyramid Extractor}
Given the moving  and  fixed images, $I_{mov}$ and $I_{fix}$, we generate two 4-level feature pyramids using siamese neural networks, that is, two neural networks with similar architecture and shared weights. Taking the moving image as an example, the feature at the $l$-th pyramid level, $f^{l}_{mov}$, is calculated by downsampling the feature at the $l-1$th level, $f^{l-1}_{mov}$ by a factor of 2 using a 3D convolutional layer with stride of 2 followed by two convoltuional layers with stride of 1. For each convolutional layer we use a Leaky ReLU as the non-linear activation function in case of sparse gradients. From the first pyramid level to the last pyramid level, the number of channels are respectively 16, 32, 64, 96.

\begin{figure}[t]
\includegraphics[width=\textwidth]{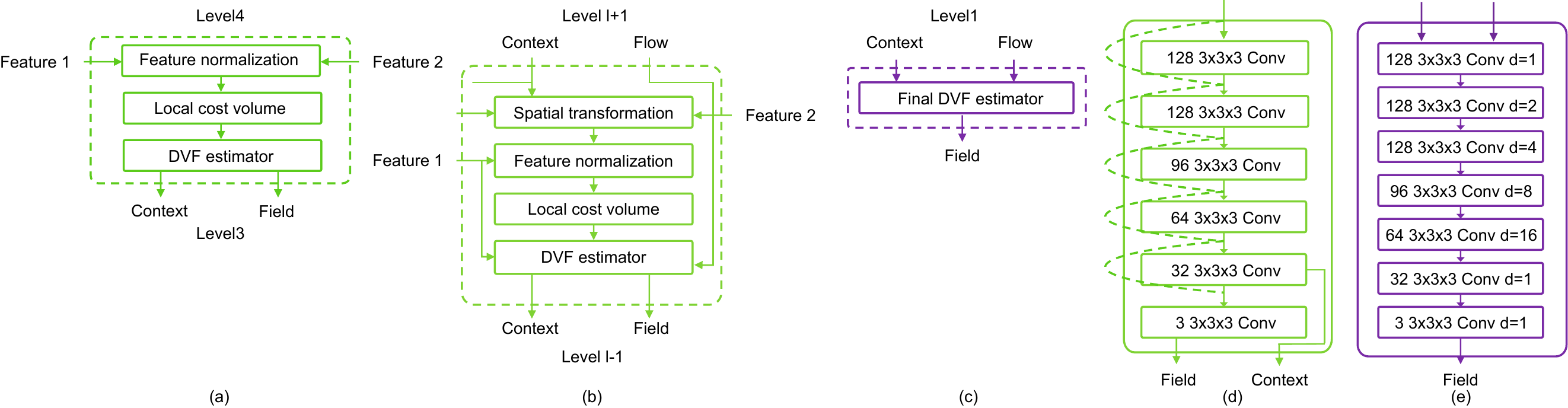}
\caption{The initial (a), intermediate (b) and final (c) deformation vector field (DVF) estimators. (d-e) are the network architectures of intermediate (d) and final (e) DVF estimators. Feature 1 represents features of the fixed image and feature 2 represents features of the moving image.} \label{fig2}
\end{figure}

\subsubsection{Feature Spatial Transformation}
At the $l$-th level (except the bottom level), features of the moving image are warped using the 2x upsampled DVF predicted at the $l+1$-th level by the spatial transformation layer\cite{jaderbergSpatialTransformerNetworks2016}:
\begin{equation}
f^{l}_{warped\_mov}=f^{l}_{mov}(x+up_2(\phi^{l+1})(x)))
\end{equation}
where $x$ is the voxel index and $up_2(\phi^{l+1})$ is the 2x upsampled DVF predicted at the $l+1$-th level. To back-propagate the gradient to both the $l$-th level features and the DVF from the last level, we use trilinear interpolation. This operation can be easily implemented by calling $\texttt{grid\_sample}$ and $\texttt{interpolate}$ functions in Pytorch.

\subsubsection{Local Cost Correlation Volume}
For 3D image registration task, constructing a cost volume for all pairs of feature vectors is prohibitive in terms of runtime and memory usage. To circumvent these limitations, we construct a local cost correlation volume to store visual similarity measures. Given the warped moving image and the fixed image features at the $l$-th level, the $l$-th local cost correlation volume is formed by taking the inner-dot product between the moving and fixed image feature vectors within a radius of $r$ units:
\begin{equation}
C(f^{l}_{warped\_mov}, f^{l}_{fix})\in \mathbb{R}^{H\times W \times D \times (2r+1)}
\end{equation}
\begin{equation}
C(x,i) = <f^{l}_{warped\_mov}(x),f^{l}_{fix}(x+i)>,\ i=-r,...,r
\end{equation}\label{EQ-CC}
where $x$ is the index of the warped image feature vectors and $x+r$ represents features vectors in the fixed image which are located within a radius $r$ of $x$ ($l_1 \ distance$). 
Compared to the concatenation of the warped features of the moving image, cost correlation volumes were found to be more discriminating for estimating DVFs. 

\subsubsection{Cost Correlation Volume Normalization}
Unsupervised similarity loss function is often only applied on the final DVF. However, several papers \cite{jonschkowskiWhatMattersUnsupervised2020,sunPWCNetCNNsOptical2018} have reported issues with feature vanishing at higher level when intermediate DVFs are not supervised.

Inspired by modern learning-based optical flow methods, we resort to cost correlation volume normalization 
at each feature level by subtracting their mean and dividing by their standard deviation before constructing cost correlation volume.

\subsubsection{Initial and Intermediate Field Estimator}
Field estimators are networks which estimate or refine the DVFs at the first or intermediate levels. At the top level, the input of the first field estimator is the cost volume and the fixed image feature at the lowest resolution. For intermediate levels, the inputs are the cost volume, the fixed image feature at current level and the upsampled DVF and context feature from the previous layer.

Each intermediate estimator has its own weights and estimation is iterated until the second level. In this paper, we chose a 5-layers dense block as the architecture of the estimator and channel outputs are 128, 128, 96, 64, 32.  

In addition to predicting a DVF, the field estimator at each level also generates contextual features that carry information across levels. In this paper, the features from  the second to last layer of the estimators are extracted as the contextual features and  used in the next level prediction.

\subsubsection{Final Field Estimator}
Traditional image registration methods often use B-spline to generate the final DVF, via interpolation of the low-resolution DVF. 
Here, we employ an individual network with a large receptive field to predict the final high-resolution DVF. 

At the last level, a 7-layers dilated convolutional network \cite{chenDeepLabSemanticImage2017} is trained with kernel size of 3x3x3. From the first to last, the channel outputs and dilation coefficients are respectively 128, 128, 128, 96, 64, 32, 3 and 1, 2, 4,8, 16, 1, 1. 

\subsection{Loss Function}
The unsupervised loss function consists  of two components: $L_{sim}$ = normalized local cross correlation, that measures the degree similarity and penalizes difference between the warped and fixed images, and $L_{reg}$= total variation (TV), that favors piecewise-smooth DVFs. $L_{reg}$ is weighted by a parameter $\lambda$.

\subsubsection{Local Cross Correlation}
Local cross correlation (LCC) is the summation of weighted cross correlation coefficients computed over patches centered at each voxel with a given window size.

\begin{equation}
LCC(I_f, I_w)= \sum_{v \in \Omega} = \frac {(\sum_{v_i} (I_f(v_i) - \bar{I_f}(v)) (I_w (v_i) - \bar{I_w} (v)) )^2} {(\sum_{v_i} (I_f(v_i)-\bar{I_f}(v))^2) (\sum_{v_i} (I_w(v_i) - \bar{I_w}(v))^2)}
\end{equation}
where $\bar{I}(v)$ is the local mean at voxel $v$, $v_i$ the index of the voxel in the patch centered at voxel $v$, $I_f$ represents the fixed image and $I_w$ is the warped image.

\subsubsection{Total Variation}
Many methods like VoxelMorph employ the smoothness to penalize large spatial variations in the deformation field. The difficulty in applying the smoothness penalty here is that there are non smooth transitions due to sliding between different parts of anatomy. To this end, TV is preferred to l2 regularisation in VoxelMorph since non smooth transition in the DVF can occur due to sliding between lung lobes:
\begin{equation}
TV(f) = \frac{1}{3 \Omega} \sum_{v\in \Omega} \sum_{i = 1,2,3} |f(v+e_i)-f(v)|
\end{equation}
where $e_i$ is the standard basis of $\mathbb{R}$. 



\section{Experiments \& Results}
\begin{figure}[t]
\includegraphics[width=\textwidth]{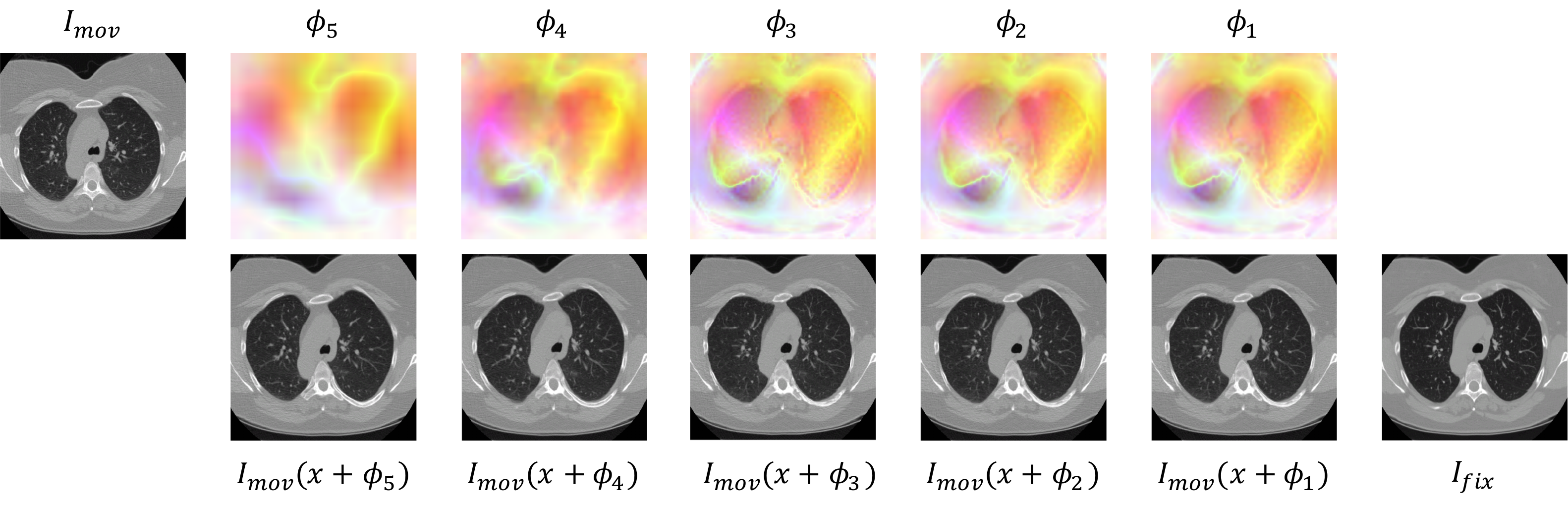}
\caption{Example of RRN registration on a pair of lung CT scans from DirLab COPDGene. The moving image ($I_{mov}$=expiratory scan) is recursively transformed by DVFs to align with the fixed image ($I_{fix}$=inspiratory scan). Each $\Phi_k$ (magnitude shown here) corresponds to the DVF predicted at the $k$th level by the DVF estimator. RGB color channels encode the DVF components in the 3 directions.}\label{fig3}
\end{figure}

\subsection{Data \& Preprocessing}
In this paper, we evaluate the performance of the RRN framework using the DirLAB COPDGene dataset. Ten cases are registered and each case contains a pair of lung CT scans acquired at full inspiration and full expiration, along with 300 annotated anatomical landmarks. After training on the ten cases using LCC and TV, we evaluated our model performance using landmarks and compared with other methods with a same metric. The average voxel size is 0.6 $\times$ 0.6 $\times$ 2.5 mm and average image dimension is 512 $\times$ 512 $\times$ 120 voxels. Based on the annotated landmarks, the range of displacement amplitude is 12-31 mm.

We first crop the scans to retain only the lung region and then resample each scan to 256 $\times$ 256 $\times$ 256 voxels. Intensities are clipped to [-1000 -100] HU. We trained the RRN over the whole dataset to register full expiration scans to their corresponding full inspiration scans. 

All models are trained with the same value of the TV weight $\lambda=0.01$. LCC and TV are applied to the DVFs generated by the final field estimator for fair comparisons. Since the performance of VoxelMorph deteriorates significantly when the batch is large, we use a batch size of 1 for all experiments of learning-based methods. We use ADAM optimizer with learning rate 1e-4, weight decay 0 and momentum 0.9 in all experiments. We stopped training our model till the value of LCC is stable. 

\subsection{Results using TRE}
\begin{table}\centering
\caption{\textbf{Comparison with state of the art classic registration methods.} Mean TRE on DirLab COPDGene dataset. The best TRE per column is highlighted in bold.}\label{tab1}
\begin{tabular}{|c|c| c| c| c| c| c| c| c| c| c |c|c|}
\hline
\multicolumn{13}{|c|}{TRE [mm]} \\
\hline
 & \#1 & \#2 & \#3 & \#4 & \#5 & \#6 & \#7 & \#8 & \#9 & \#10 & \textbf{Mean} & \textbf{Std}\\
\hline
W/o reg. &  25.90 &21.77&12.29&30.90&30.90&28.32&21.66&25.57&14.84&22.48 & 23.46 & 5.65\\
\hline
NLR \cite{ruhaakEstimationLargeMotion2017}&  1.33 & 2.34 & 1.12 & 1.54 & 1.39 & 2.08 & 1.10 & 1.57 & 0.99 & 1.42 & 1.49 & 0.39\\\hline
LMP \cite{polzinCombiningAutomaticLandmark}&  1.21 & 1.97 & 1.06 & 1.64 & 1.46 & 1.34 & 1.16 & 1.54 & 0.99 & 1.39 & 1.38 & 0.27 \\\hline
SGM3D \cite{hermannEvaluationScanLineOptimization2014}&  1.22 & 2.48 & 1.01 & 2.42 & 1.93 & 1.45 & 1.05 & 1.16 & 0.81 & 1.28 & 1.48 & 0.54\\\hline
MILO \cite{castilloComputingGlobalMinimizers2014}&  0.93 & 1.77 & 0.99 & 1.14 & 1.02 & 0.99 & 1.03 & 1.31 & 0.86 & 1.23 & 1.13 & 0.24\\\hline
MRF \cite{heinrichEstimatingLargeLung2015b}&  1.00 & 1.62 & 1.00 & 1.08 & 0.96 & 1.01 & 1.05 & 1.08 & 0.79 & 1.18 & 1.08 & 0.20\\\hline
pTV \cite{vishnevskiyIsotropicTotalVariation2017}&  \textbf{0.77} & 2.22 & \textbf{0.82} & 0.85 &\textbf{0.77} & 0.86 & \textbf{0.74} & 0.81 & 0.83 & 0.92 & 0.96 & 0.40\\\hline
\textbf{RRN} &  0.80 & \textbf{1.21} & 0.93 & \textbf{0.84} & 0.80 & \textbf{0.60} & 0.76 & \textbf{0.73} & \textbf{0.70} & \textbf{0.88} & \textbf{0.83} & \textbf{0.15}\\\hline
\end{tabular}
\end{table}
We used the target registration error (TRE) as our evaluation metric. In Table 1, we report TRE measures for our method and 6 state of the art optimization-based methods from the leaderboard. RRN best performs on 6 cases out of 10, otherwise beaten by pTV. Note that NLR, LMP, SGM3D, MILO, MRF are mask-based methods, while pTV and RRN do not use any lung masking. Since our method is optimized over the whole dataset, our method also exhibits a better standard deviation as well as the high accuracy. 

\subsection{Ablation Experiments}

We conducted a simple ablation experiment by training our model with the cost correlation volume component on and off to evaluate its impact. To be concrete, we use the concatenation of warped features and fixed features to replace the cost correlation volume. VoxelMorph is used  as the baseline in this ablation study as it does not involve pyramidal features nor cost volumes. In VoxelMorph, the moving and fixed images are directly concatenated in a two-channel 4D tensor, and  the tensor is then fed into the network. As reported in Table 2, compared to VoxelMorph, RRN drastically decreases TRE. Registration performance improves as we introduce the RRN with feature pyramids and then with cost volume. RRN is more efficient in terms of memory usage. With an input image of 256x256x256 voxels, RRN only requires about 12GB GPU memory, while VoxelMorph needs about 22GB.

Though VoxelMorph or most other encoder-decoder  unsupervised medical image registration methods and RRN are learning-based methods, the inherent principles they followed are significantly different. Unlike VoxelMorph-like methods, RNN learns two individual feature pyramids for the moving and fixed with Siamese feature extractors. This design implicitly enforces networks to encode  pairs of images to a common feature space, which is critical for downstream subnetworks to train and learn features independent of the characteristics of the moving and fixed images. In VoxelMorph, encoder features are directly fed to the decoder. Also our method exploits cost correlation volumes which we believe are more discriminative representations of the search space of DVFs and thus lead to easier learning task for CNNs. Last but not the least, our recursive refinement scheme which handles a coarse to fine registration further alleviates the difficulty of large deformation registration. 

\begin{figure}[t]
\includegraphics[width=\textwidth]{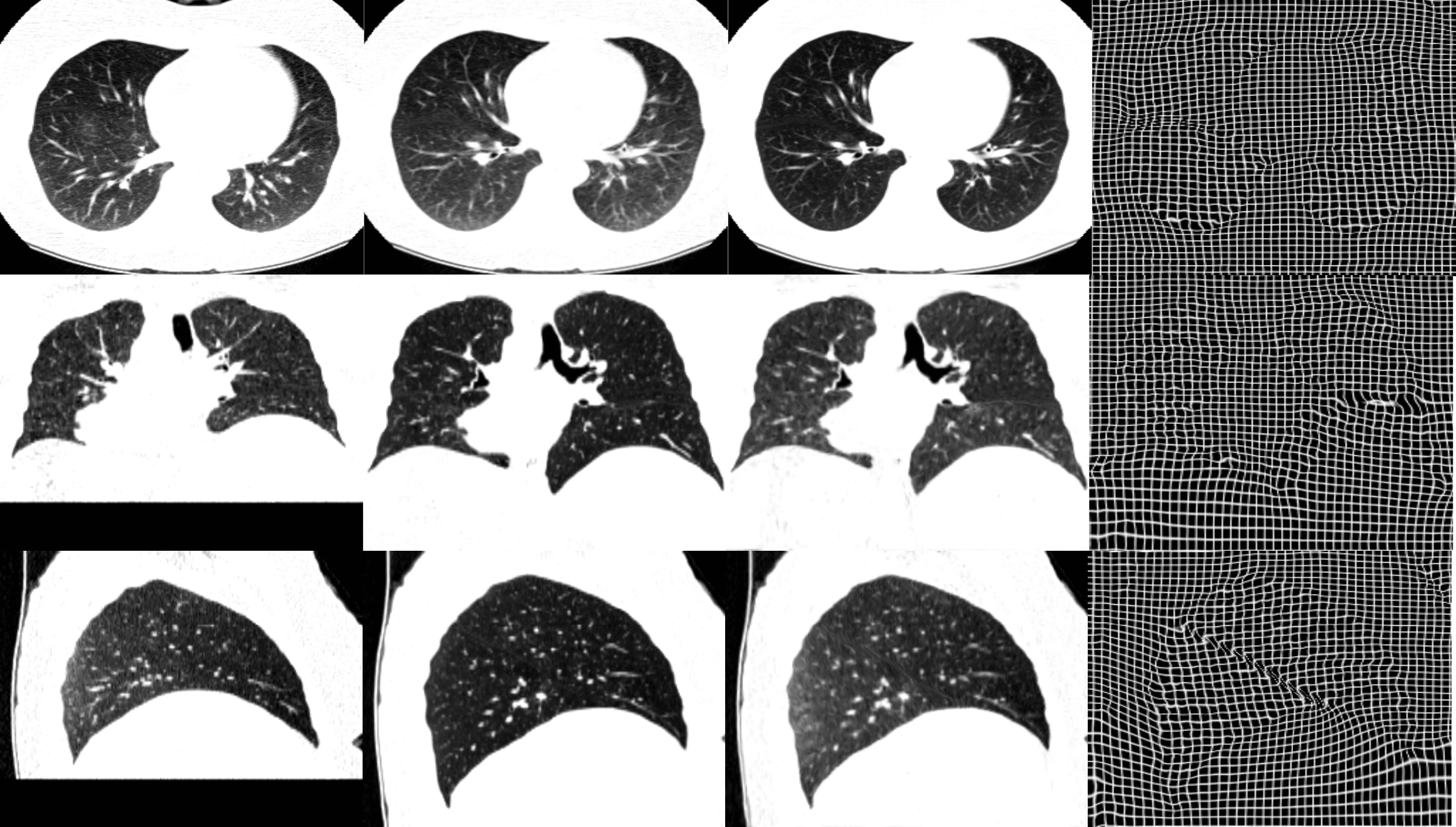}
\caption{Examples of RRN registration on a pair of lung CT scans from DirLab COPDGene. Axial, coronal and sagittal views are displayed on individual rows. From left to right: moving image, warped image, fixed image and deformation vector fields.}\label{fig4}
\end{figure}

\begin{table}\centering
\caption{ \textbf{Ablation experiment} using VoxelMorph as a benchmark learning-based method. Mean TRE are reported on the DirLab COPDGene dataset.}\label{tab2}
\begin{tabular}{|c|c| c| c| c| c| c| c| c| c| c |c|}
\hline
\multicolumn{12}{|c|}{TRE [mm]} \\
\hline
 & \#1 & \#2 & \#3 & \#4 & \#5 & \#6 & \#7 & \#8 & \#9 & \#10 & \textbf{Mean}\\
\hline
W/o reg. &  25.90 &21.77&12.29&30.90&30.90&28.32&21.66&25.57&14.84&22.48 & 23.46\\
\hline
VoxelMorph &  10.16 & 10.32 & 2.47 & 12.40 & 10.24 & 8.16 & 5.72 & 5.32 & 3.85 & 10.88 & 7.95\\\hline
RRN w/o cost volume &  0.84 & 1.56 & 0.99 & 0.92 & 0.97 & 0.68 & 0.85 & 0.84 & 0.80 & 2.53 & 1.10\\\hline
\textbf{RRN w cost volume} &  \textbf{0.80} & \textbf{1.21} & \textbf{0.93} & \textbf{0.84} & \textbf{0.80} & \textbf{0.60} & \textbf{0.76} & \textbf{0.73} & \textbf{0.70} & \textbf{0.88} & \textbf{0.83}\\\hline
\end{tabular}
\end{table}

\section{Conclusion}
 In this study, we introduced an original Recursive Refinement Network (RRN) method to perform registration of lung CT scans, using multiscale pyramidal decomposition and cost volume metrics, similarly to recent approaches for optical flow. The method was trained in an unsupervised manner. 
 RRN outperforms  state of the art pTV and VoxelMorph methods in terms of TRE on the DirLab COPDGene dataset, with an average TRE of 0.83 mm. A 13\% error reduction was obtained from the best result presented in the leaderboard of the dataset and 89\% error reduction comparing to VoxelMorph. Based on recent papers reporting TRE errors on the same data set using similar deep-learning frameworks \cite{anasCTScanRegistration2020,hansenLearningDeformablePoint2019,sokooti3DConvolutionalNeural2019a}, RRN returns the least error. RNN also has lower memory requirement than VoxelMorph. RRN is a general framework and isn't limited to a particular image type or anatomy, and could be extended to applications such as brain MRI registration or liver CT registration.

%
%
%
\bibliographystyle{splncs04}
\bibliography{rrn_v2.bib}
\end{document}